# Black phosphorus: narrow gap, wide applications

*Andres Castellanos-Gomez\*.*

Instituto Madrileño de Estudios Avanzados en Nanociencia (IMDEA-nanociencia), Campus de Cantoblanco, E-18049 Madrid, Spain.

AUTHOR INFORMATION

**Corresponding Author**

\* Andres Castellanos-Gomez (andres.castellanos@imdea.org)

The recent isolation of atomically thin black phosphorus by mechanical exfoliation of bulk layered crystals has triggered an unprecedented interest, even higher than that raised by the first works on graphene and other two-dimensional, in the nanoscience and nanotechnology community. In this Perspective we critically analyze the reasons behind the surge of experimental and theoretical works on this novel two-dimensional material. We believe that the fact that black phosphorus band gap value spans over a wide range of the electromagnetic spectrum (interesting for thermal imaging, thermoelectrics, fiber optics communication, photovoltaics, etc.) that was not covered by any other two-dimensional material isolated to date, its high carrier mobility, its ambipolar field-effect and its rather unusual in-plane anisotropy drew the attention of the scientific community towards this two-dimensional material. Here we also review the current advances, the future directions and the challenges in this young research field.

TEXT

Followed by the isolation of graphene by mechanical exfoliation of graphite,[1] many other two-dimensional materials have been produced by exfoliation of bulk layered crystals whose layers are hold together by weak van der Waals forces.[2] Up to 2013, more than 15 different two-dimensional materials were isolated by that simple yet effective fabrication method. But there was something odd however, all those two-dimensional materials (beyond graphene) were composed by two or more elements.[2,3] This statement might seem not accurate as silicene, a 2D layer made of silicon atoms, was synthesized in 2012.[4] Nonetheless silicene, unlike the other two-dimensional materials referred to above, cannot be isolated by exfoliation of bulk silicon but it is epitaxially grown onto specific substrates. Therefore, for almost 10 years people thought that graphene was a *rara avis*, the only elemental two-dimensional materials stable in free-standing form.

In 2013, a talk in the American Physical Society March meeting entitled '*Electronic Properties of Few-layer Black Phosphorus*' by Yuanbo Zhang and co-workers dethrone graphene as the only elemental two-dimensional material isolated by mechanical exfoliation.[5] According to Zhang *et al.*, atomically thin layers (and eventually one-atom thick layers) of phosphorus can be prepared by mechanical exfoliation of bulk black phosphorus (a layered allotrope of phosphorus synthesized by the first time in 1914 by Bridgman [6]). That presentation unleashed a great deal of attention on atomically thin black phosphorus, unprecedented even in the rapidly moving community of two-dimensional materials.[7–12] Indeed, according to the *ISI Web of Science* of *Thomson Reuters* in 2014 (the same year of the first papers reporting the isolation of atomically thin layers of black phosphorus) the number of publications per year on this material was already multiplied by a factor of 10 with respect to the publications per year of 2013. It took about 5 years to graphene in order to experience the same growth after its experimental isolation and molybdenum disulfide has not even reach this growth yet. But why? What makes black phosphorus more special than other recently isolated 2D materials? This Perspective analyses the facts that motivated the surge of interest on black phosphorus and critically discusses the current advances, future directions and challenges in this recent field.

**Crystal structure**

Similarly to graphite, black phosphorus atoms are strongly bonded in-plane forming layers while the layers weakly interact through van der Waals forces. But unlike in graphite (where carbon atoms bonds with three neighboring atoms through $sp^2$ hybridized orbitals) in black phosphorus the phosphorus atoms have 5 valence shell electrons available for bonding with a valence shell configuration $3s^2 3p^3$. Each phosphorus atom bonds to three neighboring phosphorus atoms through $sp^3$ hybridized orbitals, making the phosphorus atoms be arranged forming a puckered honeycomb lattice (orthorhombic, with space group *Cmca*). Each phosphorus atom also has a lone pair, which makes phosphorus very reactive to air. Figure 1a shows a schematic diagram of the black phosphorus crystal structure. This atomic arrangement yields two inequivalent directions within the black phosphorus lattice: the zigzag (parallel to the atomic ridges) and the armchair (perpendicular to the ridges). This strong structural anisotropy is the stem of its particular in-plane anisotropic electrical and optical properties that will be discussed later on in this Perspective.

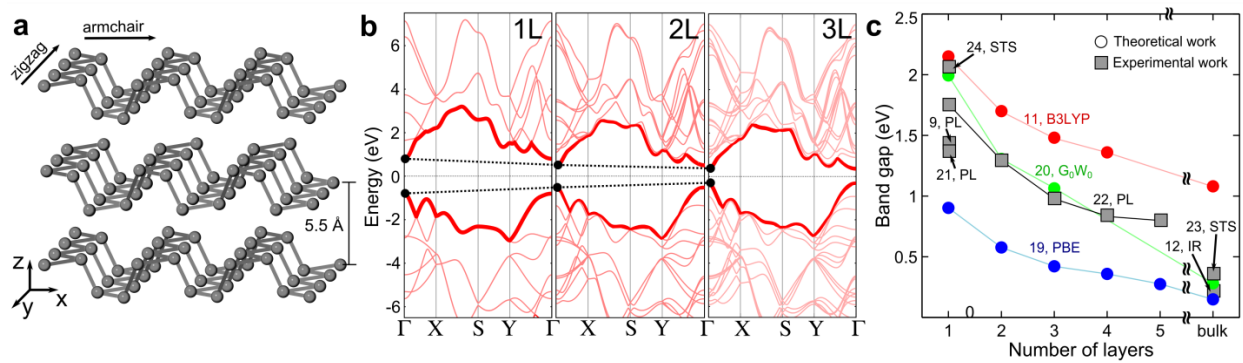

**Figure 1.** (a) Schematic diagram of the crystal structure of black phosphorus. (b) Calculated band structure of single-, bi- and tri-layer black phosphorus. Data extracted from Ref. [16]. (c) Thickness dependence of the black phosphorus band gap, calculated with different approaches. Data extracted from References [9, 11, 12, 19-24].

## Band structure: 'the right gap'

Apart from the crystal structural difference, black phosphorus also differs from graphene in the electronic band structure. While graphene is a zero-gap semiconductor (the conduction and valence bands touch at the so-called Dirac or neutrality point), black phosphorus is a semiconductor with a sizeable gap.[13-15] Figure 1b shows the calculated band structure for monolayer, bilayer and trilayer black phosphorus by *ab initio* calculations with the *GW* approximation.[16] Independently of the number of layers the band gap remains direct at the *Γ* point of the Brillouin zone. This constitutes a big difference with respect to semiconducting transition metal dichalcogenides that present a gap at the *K* point and it is only direct for single-layers.[17,18] Interestingly, the band gap value strongly depends on the number of layers. Figure 1c summarizes the thickness dependence of the black phosphorus band gap, calculated with different *ab initio* methods[10,19,20], and measured through photoluminescence [9,21,22], infrared spectroscopy [12] and scanning tunneling spectroscopy [23,24] techniques. The magnitude of the theoretical band gap strongly depends on the approximation employed to calculate the band structure, but in all the cases there is a marked thickness dependence: from a large gap (close to 2 eV) for single-layer that monotonically decreases to a narrow band gap value (about 0.3 eV) for bulk black phosphorus. This thickness dependent band gap, due to quantum confinement of the charge carriers in the out-of-plane direction, is stronger that that observed in other 2D semiconductor materials [17,18] and it provides an exceptional degree of tunability: one can select the thickness of black phosphorus to have a material with a band gap value that is optimized for a certain application. This is a very appealing feature as the band gap value spans over a wide energy range that was not covered by any of the other 2D materials isolated to date (see Figure 2). Moreover, it has been recently demonstrated that one can tune the band gap even further by alloying black phosphorus with arsenic (b-$As_xP_{1-x}$) covering the range between 0.15 eV and 0.30 eV.[25] In fact, while graphene can cover the range from 0 eV to 0.2 eV (by patterning graphene into nanoribbon shape or applying a perpendicular electric field to bilayer graphene in order to open a band gap),[26–28] transition metal dichalcogenides present a band gap in the range of 1.0-2.0 eV (depending on the thickness, strain level and chemical composition).[29,30] Thus black-phosphorus bridges the gap between graphene (a zero or nearly-zero band gap semiconductor) and transition metal dichalcogenides (wide band gap semiconductors). But what is so special about that energy range? As it can be seen at the bottom of Figure 2, a lot of applications require semiconductor materials with a band gap value in that range of energies. For example, although other two-dimensional semiconductors could be of potential use in photovoltaic energy harvesting and photocatalisys applications (which are optimized for semiconductors with 1.2 eV - 1.6 eV band gap),[31] fiber optic telecommunications (that employs wavelength in the range of 1.2 μm -1.5 μm, corresponding to photon energies of 0.8 eV - 1 eV),[32] thermal imaging (typically requires semiconductors with gaps spanning from 0.1 eV to 1.0 eV) and thermoelectric power generation (that uses materials with band gaps in the order of 0.2 eV to 0.3 eV) are out of reach of the two-dimensional semiconductors isolated to date. These applications are covered with conventional 3D semiconductors such as Pb- chalcogenides, $In_{1-x}Ga_xAs$, $In_{1-x}Ga_xSb$ and $Si_{1-x}Ge_x$. Therefore, the band gap values range spanned by black phosphorus nanolayers makes this 2D material especially suitable for thermal imaging, thermoelectric, telecom and photovoltaic applications and also makes it a prospective replacement of conventional narrow gap semiconductors in applications requiring thin, flexible and quasi-transparent material where using those 3D materials might be challenging.

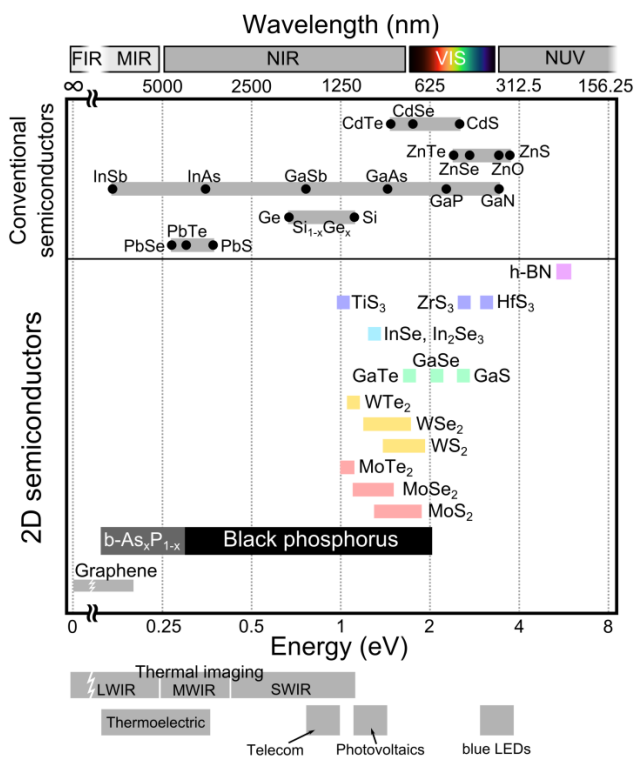

**Figure 2.** Comparison of the band gap values for different 2D semiconductor materials. The band gap values for conventional semiconductors have been also included for comparison. The horizontal bars spanning a range of band-gap values indicate that the band gap can be tuned over that range by changing the number of layers, straining or alloying. In conventional semiconductors, the bar indicates that the band gap can be continuously tuned by alloying the semiconductors (e.g. $Si_{1-x}Ge_x$ or $In_{1-x}Ga_xAs$). The range of band gap values required for certain applications have been highlighted at the bottom part of the figure to illustrate the potential applications of the different semiconductors.

## Bridging the gap between graphene and transition metal dichalcogenides

Very rapidly after the first isolating black phosphorus by mechanical exfoliation,[7–12] all the expertise build up during 10 years in graphene research was exploited to fabricate and to characterize black phosphorus based nanodevices. Figure 3 shows several examples of black phosphorus nanodevices with different functionalities. The field-effect characteristics of a black phosphorus transistor are shown in Figure 3a.[7] These kind of devices show ambipolar field-effect, high hole mobility (in the order of 10 $cm^2$/Vs - 1000 $cm^2$/Vs) and current switching ratios of 100-10000.[7–12] The ambipolarity of black phosphorus field-effect transistors is an asset in the fabrication of more complex nanodevices such as PN junctions. Note that $MoS_2$ and $WS_2$ (the most studied semiconducting transition metal dichalcogenides) tend to show a marked unipolar n-type behavior and chemical doping,[33] ionic gating [34] or specific metal contacts [35,36] have to be used to enable p-type conduction.[37,38] Recently, $WSe_2$ has also demonstrated ambipolar field effect operation.[39–42] Figure 3b shows an example where the ambipolar field-effect has been exploited to fabricate a PN junction with electrostatic split-gate geometry.[43] The generated electric field across PN junction with the two local gates can be used to separate photogenerated electron-hole pairs giving rise to a short-circuit current, a feature of photovoltaic effect in solar cells. Alternatively to the split-gate architecture, another concept of black phosphorus based photovoltaic device relies on artificial vertical stacking of two nanosheets with n-type and p-type doping to form a vertical PN junction with a built-in electric field due to the difference in doping between the two stacked layers.[44] Figure 3c shows a different example of nanodevice where the reduced dimensions and the low mass of black phosphorus nanosheets are exploited to fabricate nanoelectromechanical resonators with a natural resonance frequency in the MHz regime.[45] Coupling black phosphorus with other electronic and optoelectronic devices can be used to add new functionalities and/or to improve the performance of the fabricated black phosphorus nanodevices. Figure 3d shows an example of ultrafast photodetector fabricated by coupling a black phosphorus transistors to a silicon waveguide optical resonator.[46] To date, several photodetector devices based on black phosphorus have been demonstrated and characterized as this is a very active research area in the black phosphorus field.[11,43,44,46–52] Figure 3e displays the electrical characteristics of an inverter amplifier device fabricated by coupling an n-type $MoS_2$ transistor and a p-type black phosphorus transistor,[9] demonstrating the potential of ultrathin semiconductor materials for logic circuit applications.[9,53,54]

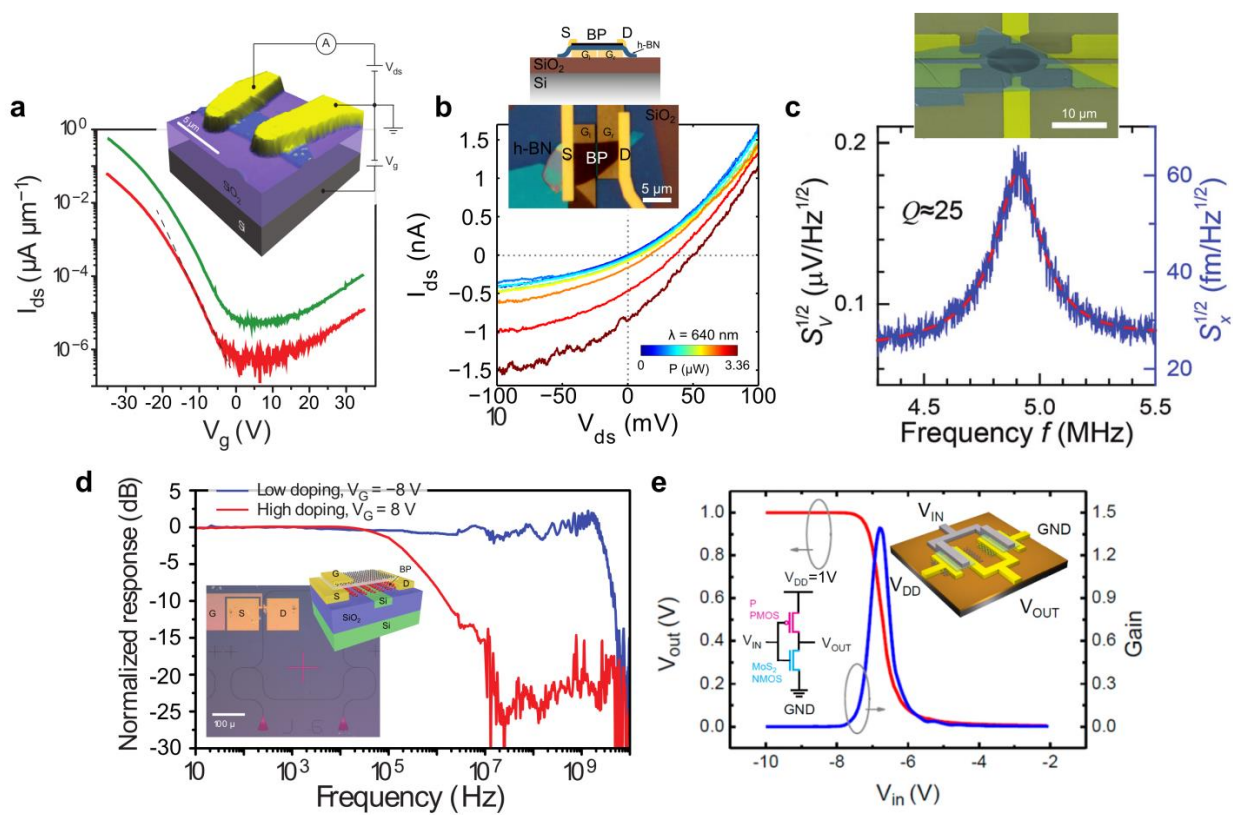

**Figure 3.** Black phosphorus based nanodevices. (a) Ambipolar field-effect transistor, adapted from Ref. [7] with permission. (b) Electrostatically-gated PN junction displaying photovoltaic effect, adapted from Ref. [43] with permission. (c) Nanoelectromechanical resonator vibrating in the very-high frequency regime, adapted from Ref. [45] with permission. (d) High frequency photodetector coupled to a silicon waveguide, adapted from Ref. [46] with permission. (e) Inverter amplifier based on $MoS_2$ and black phosphorus transistors, demonstrating the potential of black phosphorus for logic circuits, adapted from Ref. [9] with permission.

Figure 4 compares the performances of black phosphorus nanodevices with those reported for other two-dimensional materials: graphene and transition metal dichalcogenides. Figure 4a displays the charge carrier mobility *vs*. the current on/off ratio measured for field-effect transistors made out of graphene,[55–60] transition metal dichalcogenides [38,61–68] and black phosphorus.[7,8,10,11,43,47,53,54,69–73] Interestingly, the field-effect performance of black phosphorus FETs also bridges the gap between graphene (very high mobility and poor current on/off ratio) and transition metal dichalcogenides (low mobility and excellent on/off ratio). Note that the data displayed in Figure 4a is an updated version of Figure 2b from Ref. [[74]] where several recent relevant references have been added. A similar behavior is also observed for the performance of photodetectors. Figure 4b summarizes the photoresponse and the response time measured for photodetector based on graphene,[75–79] transition metal dichalcogenides [40,80–92] and black phosphorus.[11,43,46,50,51] While graphene detectors typically display a fast but rather weak photoresponse, transition metal dichalcogenides can have photoresponse values as high as $10^3$ A/W ($10^4$ times larger than that of pristine graphene based photodetectors) but they tend to suffer from very low response time, hampering their use in video imaging applications. Early black phosphorus photodetectors already showed photoresponse times faster than most of transition metal dichalcogenide based photodetectors.[11,43,50,51] Note that the time response values reported for these early black phosphorus photodetectors are already very close to *RC* time of the experimental setup (~1 ms) and thus these values can be considered as lower bounds for the black phosphorus time response. In fact, more sophisticated device engineering (integrating a black phosphorus detector on a silicon photonic waveguide to overcome the *RC* time limitation in the photodetection setup) allowed to demonstrate a time response of a black phosphorus photodetector

< 1 ns.[46] The data displayed in Figure 4b has been adapted from Figure 12a in Ref. [[49]], including some recent relevant references. Finally, Figure 4 also presents a comparison between two of the most important physical properties for thermoelectric power generation applications: the Seebeck coefficient and the electrical resistivity. An ideal thermoelectric material should present a high Seebeck coefficient and a low electrical resistivity. Note that the amount of reported data for thermoelectric devices is still very limited (and mostly limited to theoretical studies) in comparison with electronic and optoelectronic devices.[93–99] According to Figure 4c, transition metal dichalcogenides can reach sky high values of the Seebeck coefficient but with high electrical resistivity and graphene, on the other hand, presents rather low Seebeck coefficient but with a small electrical resistivity.[100–103] For black phosphorus, the only direct measurement of the thermoelectric power has been done for bulk samples finding a Seebeck coefficient value of +345 μV/K with an electrical resistance in between that of graphene and transition metal dichalcogenides.[98] Figure 4c also represents theoretical data calculated for black phosphorus nanolayers with different density of charge carriers,[94,95,99] which shows that using electrostatic field-effect one could substantially reduce electrical resistance while keeping a high enough Seebeck coefficient which motivates the exploration of black phosphorus as a thermoelectric material.

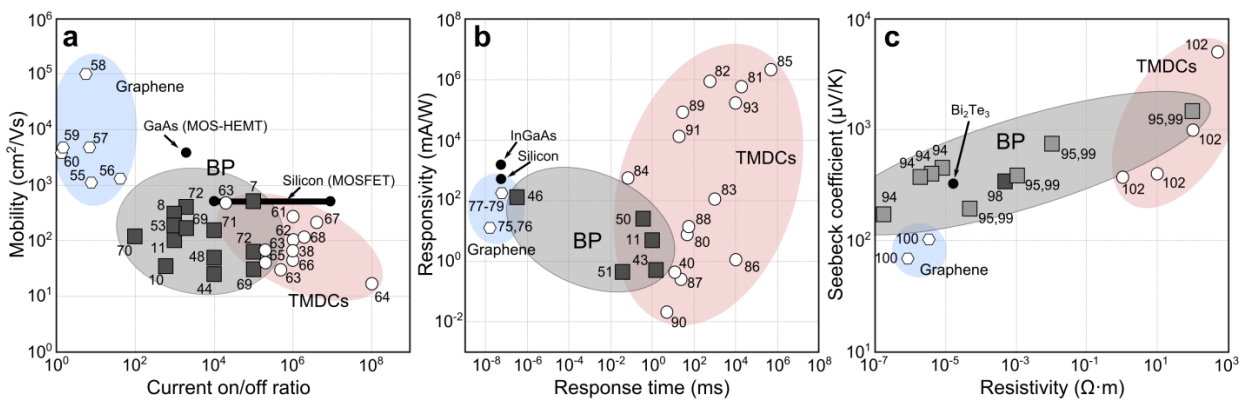

**Figure 4.** Comparison between the performances of 2D-based nanodevices. (a) Carrier mobility *vs.* current on/off ratio reported for field-effect transistors based on graphene, TMDCs and black phosphorus. (b) Summary of the responsivity and time response of photodetectors based on graphene, TMDCs and black phosphorus. The data displayed in (a) and (b) are updated versions of Figure 2b in Ref. [74] and Figure 12a in Ref. [49], including recent relevant references. (c) Seebeck coefficient *vs.* electrical responsivity of graphene, TMDCs and black phosphorus.

## Rather unusual in-plane anisotropy

As discussed above, the crystal structure of black phosphorus (sketched in Figure 1a) is the stem of its in-plane anisotropic properties.[7,9,12,19–21,47,51,96,97,104–122] Graphene, boron nitride or Mo- and W- based transition metal dichalcogenides, on the other hand, do not present noticeable in-plane anisotropy. Figure 5 summarizes the in-plane angular distribution of electrical, optical and mechanical properties of black phosphorus. Field-effect devices with electrodes arranged at different angles have been used to probe the anisotropy of the charge carrier mobility and the electrical conductance, finding a higher mobility and conductance along the armchair direction (up to 50% higher than along the zigzag direction at room temperature).[7,9,12,113] Figure 5a shows a polar plot with the angular dependence of the DC conductivity measured in a black phosphorus device with electrodes patterned with an angular spacing of 30º (see the inset).[12]The relative optical extinction, measured on the black phosphorus device in Figure 5a, is shown in Figure 5b for different polarization angles (relative to the horizontal axis) showing an even stronger anisotropy.[12] This marked linear dichroism of black phosphorus, optical absorption that depends on the relative orientation between the materials lattice and an incident linearly polarized light, has also strong implications in its Raman spectra, plasmonic and screening effects, photoresponse and photoluminescence emission yield.[20,21,47,51,113,120,123,124] For example, Figure 5c shows a polar plot with the polarization angle of the photoluminescence yield of a black phosphorus flake excited with a laser linearly polarized along the zigzag direction, armchair direction and a direction at 45º between the zigzag and armchair directions. For all the different excitation laser polarization angles, the photoluminescence yield always shows a high degree of polarization along the armchair direction of the flake.[21] Finally, Figure 5d shows the calculated angular dependence of the Young's modulus of black phosphorus which is predicted to be up to 4 times higher along the zigzag direction ($E_{zz}$ = 166 GPa) than along the armchair ($E_{ac}$ = 44 GPa),[122,125,126] although no direct experimental evidence of this anisotropic mechanical properties has been reported yet. This characteristic in-plane anisotropy of black phosphorus can be exploited to study quasi-1D physics in a system with 2D geometry, which is easier to integrate with other devices and to interact with them because of their large surface area. Moreover, the in-plane anisotropy can also provide with novel functionality to devices based on black phosphorus (see the Outlook and Summary section).

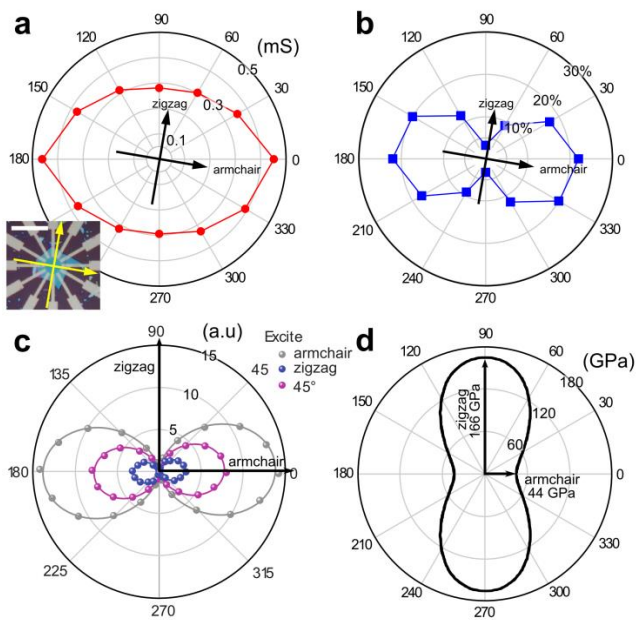

**Figure 5.** In-plane anisotropic properties of black phosphorus. (a) Angular dependence of the DC electrical conductance of a black phosphorus transistor. (b) Angular dependence of the relative optical extinction measured on the same flake displayed in (a). Panels (a) and (b) have been adapted from Ref. [12] with permission. (c) Angular dependent photoluminescence yield of a black phosphorus flake with laser excitation ($\lambda$ = 532 nm) polarized along the zigzag direction, armchair direction and a 45º direction. Adapted from Ref. [21] with permission. (d) Calculated angular dependence of the black phosphorus Young's modulus. The calculation predicts that the zigzag direction is about 4 times stiffer than the armchair direction. Adapted from Ref. [122] with permission.

## Challenges: isolation and stability

In the following, we will discuss the main two challenges that the black phosphorus research is facing so far: the fabrication of large area samples and the environmental instability of black phosphorus. Most of the experiments reported on black phosphorus in the literature relied on mechanically exfoliated samples. This fabrication method, although it provides high quality material for research, it is not scalable in industrial applications. Recently demonstrated liquid-phase exfoliation, on the other hand, offers a cheap and scalable alternative to isolate ultrathin black phosphorus layers but the quality of the produced material might not be high enough for certain electronic or optical applications.[127–132] Very recently, another alternative method to fabricate black phosphorus thin films based on pulsed laser deposition, a technology that can be scaled up, has been presented.[133] The fabricated layers, however, lack of crystallinity and their electrical properties are still way poorer that those of pristine black phosphorus material. Therefore in order to transfer the black phosphorus from research laboratories to the first real-life applications one should solve the challenge of developing a large area synthesis method to fabricate highly crystalline black phosphorus layers. For graphene, it took 5 years since the first mechanical exfoliation experiments till the development of the first chemical vapor deposition recipes to fabricate large area single-layer graphene.[134,135] This time was significantly reduced to about 2 years for transition metal dichalcogenides,[136,137] indicating that the scientific community intensified the effort to develop large-area fabrication techniques to grow high quality wafer-scale two-dimensional materials. Thus the fabrication of large area black phosphorus nanolayers is surely motivating the groups specialized on synthesis of nanomaterials nowadays as a reliable recipe to fabricate large area crystalline black phosphorus thin films will undoubtedly have a strong impact in the two-dimensional materials community.

Another current challenge in the black phosphorus research community is the environmental-induced degradation of the black phosphorus.[8,10,138–141] Most of the two-dimensional materials studied to date (graphene, h-BN, $MoS_2$, etc.) are rather stable at atmospheric conditions which facilitated the fabrication of devices based on these materials. Back phosphorus, however, shows a relatively large reactivity.[142,143] Exfoliated flakes of black phosphorus are highly hygroscopic and tend to uptake moisture from air (see Figure 6a).[138] The long term contact with the water condensed on the surface seems to degrade the black phosphorus, as it can be seen by measuring the electrical performance of transistors [138,139] or the sheet resistance (measured with an atomic force microscopy based microwave impedance microscope) [140] as a function of time (Figure 6b and 6c). In the work by Favron *et al*.[141] a systematic study of the degradation mechanisms in black phosphorus, combining Raman and transmission electron spectroscopies, allowed to determine that the most plausible mechanism behind the aging of black phosphorus in air is a photoassisted oxidation reaction with oxygen dissolved in the adsorbed on its surface. Nonetheless, the timescale at which noticeable degradation takes place is long enough to allow the fabrication of black phosphorus devices but they should be operated in vacuum conditions to ensure their reliability. A different approach relies on encapsulating the black phosphorus devices at an early stage during the fabrication.[70,139,140] Different encapsulation options have been explored so far: polymer dielectrics,[50,144] atomic layer deposited oxides [54,69,139,140] and van der Waals heterostructures with boron nitride flakes.[70,73,145–149] Figure 6b and 6c show the time evolution of the electrical properties of black phosphorus flakes encapsulated with an aluminum oxide layer.[139,140] While the pristine, unencapsulated, flakes degrade in a time scale of 1-2 days, the encapsulated flakes remains unaltered for several days. Figure 6d shows the fabrication steps necessary to encapsulate black phosphorus flakes between two dielectric flakes of hexagonal boron nitride and to make electrical contacts to the edges of black phosphorus.[148] This encapsulation and contacting method has been recently developed to fabricate high electronic quality graphene and $MoS_2$ samples, which showed record-high charge carrier mobilities.[150,151] The black phosphorus devices fabricated by this method also present charge carrier mobility values higher than conventional black phosphorus field-effect transistors and their excellent electronic properties have been probed by studying quantum oscillations at low temperatures.[145–149] Therefore, despite of the youth of the black phosphorus research, the experimental efforts to overcome its environmental instability by encapsulation are at a very advanced stage.

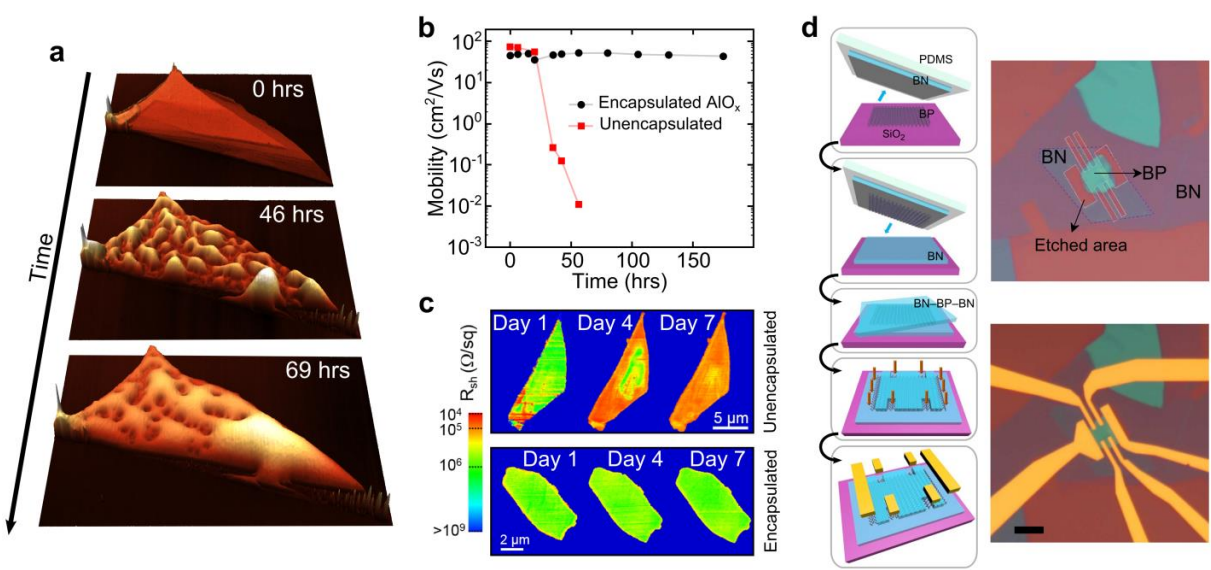

**Figure 6.** Environmental-induced degradation of black phosphorus and its encapsulation as a route to preserve its pristine properties. (a) Atomic force microscopy topography images acquired as a function of time. Due to the high hygroscopic character of black phosphorus it uptakes moisture from air. Adapted from Ref. [138] with permission. (b) Time evolution of the charge carrier mobility of black phosphorus field-effect transistors. Unencapsulated devices rapidly degrade upon air exposure while devices encapsulated with an aluminum oxide layer shows almost time independent performance for several days. Adapted from Ref. [139] with permission. (c) Microwave impedance microscopy map of the local sheet resistance of two black phosphorus flakes: one unencapsulated and one encapsulated with a thick film of aluminum oxide. While the encapsulated device remains unaltered for several days, the unencapsulated device resistance strongly changes with time, starting at the edges of the flake and growing towards the center. Adapted from Ref. [140] with permission. (d) Sandwiching of black phosphorus between h-BN fakes to fabricate high-quality encapsulated devices with edge electrical contacts. Adapted from Ref. [148] with permission.

**Outlook and summary**

In the last part of this Perspective, we will briefly discuss the possible future avenues in the black phosphorus research. One can anticipate that among the most urgent topics one will be, as previously discussed in the paragraph devoted to the challenges, the development of a large-area fabrication technique to isolate high quality wafer scale thin films of black phosphorus as that result could have a big impact not only in the scientific community but also in the first applications based on black phosphorus. The further development of experimental techniques to reliably determine the thickness of black phosphorus [152–155] and to characterize its electrical, optical and chemical properties will also take place in the first years of black phosphorus research.[24,156–158]

Because of its narrow band gap value, multilayered black phosphorus will also find a place in photodetection in the near- and mid-infrared part of the spectrum and in thermoelectric power generation applications. Although some black phosphorus photodetector devices have been already demonstrated and characterized, only simple architectures have been studied so far and there is plenty of room to optimize the performance of the devices, to explore unconventional device architectures and to combine black phosphorus with other two-dimensional materials or other semiconductor materials.[44,159–162] For instance, the combination of graphene with semiconducting quantum dots have demonstrated to improve the detectivity of graphene-based photodetectors by a factor of $10^9$.[163] Within the field of optoelectronics, other topic that has attracted considerable interest is the use of two dimensional semiconductors for electrically generated light emitting applications (electroluminescence). Recent works demonstrated that $MoS_2$,[164] $WS_2$,[165] $WSe_2$ [39,41,42] and artificial heterostructures [166] can be used as atomically thin light emitting devices, generating photons in the range of 1.5 eV to 2.0 eV. Black phosphorus light emitting devices hold the promise to emit photons in the NIR part of the spectrum.

Regarding the application of black phosphorus on thermoelectric applications, despite of the recent theoretical interest on the thermoelectric properties of black phosphorus [93–97,99] the amount of experimental works is still very scarce and it is basically a completely open field that will bloom in the near future because of the promising properties of this material for thermoelectrics. For instance the inherent in-plane anisotropic properties of black phosphorus can be exploited to increase the performance of thermoelectric power generation devices as the high electrical conductance and with high thermal conductance crystal directions are orthogonal.[99] Therefore, in black phosphorus one can engineer a device with low electrical resistance that can keep a large thermal gradient along the device, which are the requirements for a highly efficient thermopower generator.

Another interesting aspect in black phosphorus research is the use of strain engineering, the modification of the electronic and optical properties of a material by subjecting it to a mechanical deformation. In fact, strain engineering in 2D materials,[167] not only in black phosphorus, is recently gaining more and more attention as they can typically stand very large deformations before the rupture (around 10%) and thus one can widely tune the electrical/optical properties through strain engineering.[168–174] Conventional 3D semiconductors, on the contrary, tend to break for moderate deformations (<1.5%). For black phosphorus, strain engineering might be even more interesting than for other 2D semiconductors because of its in-plane anisotropy.[94,96,105,106,119,175–177] Recent theoretical works predicted that one can control the preferable direction for charge carriers by changing the level of strain on the black phosphorus lattice.[106] Nonetheless, the amount of works on strain engineering in black phosphorus is still limited and mainly restricted to theoretical studies.[178]

In summary, this Perspective article has reviewed the recent advances in the research on black phosphorus. The aspects that motivated the large surge of experimental and theoretical works on this recently isolated material has been critically discussed finding several factors that justify the recent interest on this material: (1) its band gap spans over a range of the spectrum not covered by any other 2D material isolated to date, (2) it presents a good trade-off between charge carrier mobility and current on/off switching ratios which makes it a prospective material for digital and high frequency electronics, (3) it shows a rather exotic in-plane anisotropy that can be exploited to engineer new concepts of devices. Another remarkable characteristic of this material is that black phosphorus electronic, optoelectronic and thermoelectric nanodevices display performances that bridge those of graphene and transition metal dichalcogenides. This trade-off between graphene-like (rather metallic) and transition metal dichalcogenide-like (wide band gap semiconductor) can be optimal (or even crucial) for many different applications such as thermal imaging, photodetection in the telecom range or thermopower generation from thermal waste. The challenges and the open questions in this young research field have been also discussed in the last part of this Perspective. We consider that the two main challenges are (1) the fabrication of wafer-scale crystalline black phosphorus thin films and (2) overcoming the environmental induced degradation of black phosphorus by finding encapsulation methods that preserve its pristine properties at long-term. Among the possible future directions of the field, we highlighted here (1) the integration of black phosphorus in complex device architectures or combining it with other nanomaterials to enhance the performance, (2) the experimental exploration of black phosphorus as a thermoelectric material (especially exploiting its anisotropy to decouple the high electrical and thermal conductance directions) and (3) the use of strain engineering to control the electrical and optical properties of black phosphorus (and even to control the preferential in-plane conduction directions for charge carriers). As a final remark, due to the fast growing rate of the community studying this material and the amount of exciting new avenues just opened to date the author is confident that most of the breakthroughs in this field are still to come.

*Note added*: During the reviewing process of this manuscript the author became aware of the publication of a work by Xuesong Li *et al.* entitled "Synthesis of thin-film black phosphorus on a flexible substrate" where the authors demonstrate the synthesis of a large area thin-film of black phosphorus (4 mm$^2$ and 40 nm thick).[179]

AUTHOR INFORMATION

**Corresponding Author:** Andres Castellanos-Gomez

*Email: andres.castellanos@imdea.org

ACKNOWLEDGMENT

AC-G acknowledges financial support from the BBVA Foundation through the fellowship "I

Convocatoria de Ayudas Fundacion BBVA a Investigadores, Innovadores y Creadores

Culturales" ("Semiconductores ultradelgados: hacia la optoelectronica flexible").

REFERENCES

(1)    Novoselov, K. S.; Geim, a K.; Morozov, S. V; Jiang, D.; Zhang, Y.; Dubonos, S. V;
       Grigorieva, I. V; Firsov, A. A. Electric Field Effect in Atomically Thin Carbon Films.
       *Science* **2004**, *306* (5696), 666–669.

(2)    Novoselov, K. S.; Jiang, D.; Schedin, F.; Booth, T. J.; Khotkevich, V. V; Morozov, S. V;
       Geim, a K. Two-Dimensional Atomic Crystals. *Proc. Natl. Acad. Sci. U. S. A.* **2005**, *102*
       (30), 10451–10453.

(3)    Coleman, J. N.; Lotya, M.; O'Neill, A.; Bergin, S. D.; King, P. J.; Khan, U.; Young, K.;
       Gaucher, A.; De, S.; Smith, R. J.; et al. Two-Dimensional Nanosheets Produced by Liquid
       Exfoliation of Layered Materials. *Science* **2011**, *331* (6017), 568–571.

(4)    Vogt, P.; De Padova, P.; Quaresima, C.; Avila, J.; Frantzeskakis, E.; Asensio, M. C.;
       Resta, A.; Ealet, B.; Le Lay, G. Silicene: Compelling Experimental Evidence for
       Graphenelike Two-Dimensional Silicon. *Phys. Rev. Lett.* **2012**, *108* (15), 155501.

(5)    Li, L.; Yu, Y.; Ye, G. J.; Chen, X. H.; Zhang, Y. Electronic Properties of Few-Layer
       Black Phosphorus. *Am. Phys. Soc.* **2013**.

(6)    Bridgman, P. W. Two New Modifications of Phosphorus. *J. Am. Chem. Soc.* **1914**, *36* (7),
       1344–1363.

(7)    Li, L.; Yu, Y.; Ye, G. J.; Ge, Q.; Ou, X.; Wu, H.; Feng, D.; Chen, X. H.; Zhang, Y. Black
       Phosphorus Field-Effect Transistors. *Nat. Nanotechnol.* **2014**, *9* (5), 372–377.

(8)     Koenig, S. P.; Doganov, R. A.; Schmidt, H.; Castro Neto, A. H.; Özyilmaz, B. Electric Field Effect in Ultrathin Black Phosphorus. *Appl. Phys. Lett.* **2014**, *104* (10), 103106.

(9)     Liu, H.; Neal, A. T.; Zhu, Z.; Luo, Z.; Xu, X.; Tománek, D.; Ye, P. D. Phosphorene: An Unexplored 2D Semiconductor with a High Hole Mobility. *ACS Nano* **2014**, *8* (4), 4033–4041.

(10)    Castellanos-Gomez, A.; Vicarelli, L.; Prada, E.; Island, J. O.; Narasimha-Acharya, K. L.; Blanter, S. I.; Groenendijk, D. J.; Buscema, M.; Steele, G. A.; Alvarez, J. V.; et al. Isolation and Characterization of Few-Layer Black Phosphorus. *2D Mater.* **2014**, *1* (2), 025001.

(11)    Buscema, M.; Groenendijk, D. J.; Blanter, S. I.; Steele, G. A.; van der Zant, H. S. J.; Castellanos-Gomez, A. Fast and Broadband Photoresponse of Few-Layer Black Phosphorus Field-Effect Transistors. *Nano Lett.* **2014**, *14* (6), 3347–3352.

(12)    Xia, F.; Wang, H.; Jia, Y. Rediscovering Black Phosphorus as an Anisotropic Layered Material for Optoelectronics and Electronics. *Nat. Commun.* **2014**, *5*, 4458.

(13)    Keyes, R. W. The Electrical Properties of Black Phosphorus. *Phys. Rev.* **1953**, *92* (3), 580–584.

(14)    Asahina, H.; Morita, A. Band Structure and Optical Properties of Black Phosphorus. *J. Phys. C Solid State Phys.* **1984**, *17* (11), 1839–1852.

(15)    Morita, A. Semiconducting Black Phosphorus. *Appl. Phys. A Solids Surfaces* **1986**, *39* (4), 227–242.

(16)    Rudenko, A. N.; Katsnelson, M. I. Quasiparticle Band Structure and Tight-Binding Model for Single- and Bilayer Black Phosphorus. *Phys. Rev. B* **2014**, *89* (20), 201408.

(17)    Splendiani, A.; Sun, L.; Zhang, Y.; Li, T.; Kim, J.; Chim, C.-Y.; Galli, G.; Wang, F. Emerging Photoluminescence in Monolayer MoS2. *Nano Lett.* **2010**, *10* (4), 1271–1275.

(18)    Mak, K. F.; Lee, C.; Hone, J.; Shan, J.; Heinz, T. F. Atomically Thin MoS_{2}: A New Direct-Gap Semiconductor. *Phys. Rev. Lett.* **2010**, *105* (13), 136805.

(19)    Qiao, J.; Kong, X.; Hu, Z.-X.; Yang, F.; Ji, W. High-Mobility Transport Anisotropy and Linear Dichroism in Few-Layer Black Phosphorus. *Nat. Commun.* **2014**, *5*, 4475.

(20)    Tran, V.; Soklaski, R.; Liang, Y.; Yang, L. Layer-Controlled Band Gap and Anisotropic Excitons in Few-Layer Black Phosphorus. *Phys. Rev. B* **2014**, *89* (23), 235319.

(21)  Wang, X.; Jones, A. M.; Seyler, K. L.; Tran, V.; Jia, Y.; Zhao, H.; Wang, H.; Yang, L.; Xu, X.; Xia, F. Highly Anisotropic and Robust Excitons in Monolayer Black Phosphorus. *Nat. Nanotechnol.* **2015**, *10* (6), 517–521.

(22)  Yang, J.; Xu, R.; Pei, J.; Myint, Y. W.; Wang, F.; Wang, Z.; Zhang, S.; Yu, Z.; Lu, Y. Optical Tuning of Exciton and Trion Emissions in Monolayer Phosphorene. *Light Sci. Appl.* **2015**, *4* (7), e312.

(23)  Zhang, C. D.; Lian, J. C.; Yi, W.; Jiang, Y. H.; Liu, L. W.; Hu, H.; Xiao, W. D.; Du, S. X.; Sun, L. L.; Gao, H. J. Surface Structures of Black Phosphorus Investigated with Scanning Tunneling Microscopy. *J. Phys. Chem. C* **2009**, *113* (43), 18823–18826.

(24)  Liang, L.; Wang, J.; Lin, W.; Sumpter, B. G.; Meunier, V.; Pan, M. Electronic Bandgap and Edge Reconstruction in Phosphorene Materials. *Nano Lett.* **2014**, *14* (11), 6400–6406.

(25)  Liu, B.; Köpf, M.; Abbas, A. N.; Wang, X.; Guo, Q.; Jia, Y.; Xia, F.; Weihrich, R.; Bachhuber, F.; Pielnhofer, F.; et al. Black Arsenic-Phosphorus: Layered Anisotropic Infrared Semiconductors with Highly Tunable Compositions and Properties. *Adv. Mater.* **2015**, *27* (30), 4423–4429.

(26)  Chen, Z.; Lin, Y.-M.; Rooks, M. J.; Avouris, P. Graphene Nano-Ribbon Electronics. *Phys. E Low-dimensional Syst. Nanostructures* **2007**, *40* (2), 228–232.

(27)  Castro, E.; Novoselov, K.; Morozov, S.; Peres, N.; dos Santos, J.; Nilsson, J.; Guinea, F.; Geim, A.; Neto, A. Biased Bilayer Graphene: Semiconductor with a Gap Tunable by the Electric Field Effect. *Phys. Rev. Lett.* **2007**, *99* (21), 216802.

(28)  Oostinga, J. B.; Heersche, H. B.; Liu, X.; Morpurgo, A. F.; Vandersypen, L. M. K. Gate-Induced Insulating State in Bilayer Graphene Devices. *Nat. Mater.* **2008**, *7* (2), 151–157.

(29)  Wang, Q. H.; Kalantar-Zadeh, K.; Kis, A.; Coleman, J. N.; Strano, M. S. Electronics and Optoelectronics of Two-Dimensional Transition Metal Dichalcogenides. *Nat. Nanotechnol.* **2012**, *7* (11), 699–712.

(30)  Lv, R.; Robinson, J. A.; Schaak, R. E.; Sun, D.; Sun, Y.; Mallouk, T. E.; Terrones, M. Transition Metal Dichalcogenides and Beyond: Synthesis, Properties, and Applications of Single- and Few-Layer Nanosheets. *Acc. Chem. Res.* **2015**, *48* (1), 56–64.

(31)  Shockley, W.; Queisser, H. J. Detailed Balance Limit of Efficiency of P-N junction Solar Cells. *J. Appl. Phys.* **1961**, *32* (3), 510.

(32)  Soole, J. B. D.; Schumacher, H. InGaAs Metal-Semiconductor-Metal Photodetectors for Long Wavelength Optical Communications. *IEEE J. Quantum Electron.* **1991**, *27* (3), 737–752.

(33)   Laskar, M. R.; Nath, D. N.; Ma, L.; Lee, E. W.; Lee, C. H.; Kent, T.; Yang, Z.; Mishra, R.; Roldan, M. A.; Idrobo, J.-C.; et al. P-Type Doping of MoS2 Thin Films Using Nb. *Appl. Phys. Lett.* **2014**, *104* (9), 092104.

(34)   Zhang, Y.; Ye, J.; Matsuhashi, Y.; Iwasa, Y. Ambipolar MoS2 Thin Flake Transistors. *Nano Lett.* **2012**, *12* (3), 1136–1140.

(35)   Fontana, M.; Deppe, T.; Boyd, A. K.; Rinzan, M.; Liu, A. Y.; Paranjape, M.; Barbara, P. Electron-Hole Transport and Photovoltaic Effect in Gated MoS2 Schottky Junctions. *Sci. Rep.* **2013**, *3*, 1634.

(36)   Chuang, S.; Battaglia, C.; Azcatl, A.; McDonnell, S.; Kang, J. S.; Yin, X.; Tosun, M.; Kapadia, R.; Fang, H.; Wallace, R. M.; et al. MoS$_2$ P-Type Transistors and Diodes Enabled by High Work Function MoOx Contacts. *Nano Lett.* **2014**, *14* (3), 1337–1342.

(37)   Radisavljevic, B.; Radenovic, A.; Brivio, J.; Giacometti, V.; Kis, A. Single-Layer MoS2 Transistors. *Nat. Nanotechnol.* **2011**, *6* (3), 147–150.

(38)   Ovchinnikov, D.; Allain, A.; Huang, Y.-S.; Dumcenco, D.; Kis, A. Electrical Transport Properties of Single-Layer WS2. *ACS Nano* **2014**, *8* (8), 8174–8181.

(39)   Ross, J. S.; Klement, P.; Jones, A. M.; Ghimire, N. J.; Yan, J.; Mandrus, D. G.; Taniguchi, T.; Watanabe, K.; Kitamura, K.; Yao, W.; et al. Electrically Tunable Excitonic Light-Emitting Diodes Based on Monolayer WSe2 P-N Junctions. *Nat. Nanotechnol.* **2014**, *9* (4), 268–272.

(40)   Groenendijk, D. J.; Buscema, M.; Steele, G. A.; Michaelis de Vasconcellos, S.; Bratschitsch, R.; van der Zant, H. S. J.; Castellanos-Gomez, A. Photovoltaic and Photothermoelectric Effect in a Double-Gated WSe2 Device. *Nano Lett.* **2014**, *14* (10), 5846–5852.

(41)   Baugher, B. W. H.; Churchill, H. O. H.; Yang, Y.; Jarillo-Herrero, P. Optoelectronic Devices Based on Electrically Tunable P-N Diodes in a Monolayer Dichalcogenide. *Nat. Nanotechnol.* **2014**, *9* (4), 262–267.

(42)   Pospischil, A.; Furchi, M. M.; Mueller, T. Solar-Energy Conversion and Light Emission in an Atomic Monolayer P-N Diode. *Nat. Nanotechnol.* **2014**, *9* (4), 257–261.

(43)   Buscema, M.; Groenendijk, D. J.; Steele, G. A.; van der Zant, H. S. J.; Castellanos-Gomez, A. Photovoltaic Effect in Few-Layer Black Phosphorus PN Junctions Defined by Local Electrostatic Gating. *Nat. Commun.* **2014**, *5*, 4651.

(44) Deng, Y.; Luo, Z.; Conrad, N. J.; Liu, H.; Gong, Y.; Najmaei, S.; Ajayan, P. M.; Lou, J.; Xu, X.; Ye, P. D. Black Phosphorus-Monolayer MoS2 van Der Waals Heterojunction P-N Diode. *ACS Nano* **2014**, *8* (8), 8292–8299.

(45) Wang, Z.; Jia, H.; Zheng, X.; Yang, R.; Wang, Z.; Ye, G.; X. H., C.; Shan, J.; Feng, P. Black Phosphorus Nanoelectromechanical Resonators Vibrating at Very High Frequencies. *Nanoscale* **2015**, *7*, 877–884.

(46) Youngblood, N.; Chen, C.; Koester, S. J.; Li, M. Waveguide-Integrated Black Phosphorus Photodetector with High Responsivity and Low Dark Current. *Nat. Photonics* **2015**, *9* (4), 247–252.

(47) Hong, T.; Chamlagain, B.; Lin, W.; Chuang, H.-J.; Pan, M.; Zhou, Z.; Xu, Y.-Q. Polarized Photocurrent Response in Black Phosphorus Field-Effect Transistors. *Nanoscale* **2014**, *6* (15), 8978–8983.

(48) Low, T.; Engel, M.; Steiner, M.; Avouris, P. Origin of Photoresponse in Black Phosphorus Phototransistors. *Phys. Rev. B* **2014**, *90* (8), 081408.

(49) Buscema, M.; Island, J. O.; Groenendijk, D. J.; Blanter, S. I.; Steele, G. A.; van der Zant, H. S. J.; Castellanos-Gomez, A. Photocurrent Generation with Two-Dimensional van Der Waals Semiconductors. *Chem. Soc. Rev.* **2015**, *44* (11), 3691–3718.

(50) Engel, M.; Steiner, M.; Avouris, P. Black Phosphorus Photodetector for Multispectral, High-Resolution Imaging. *Nano Lett.* **2014**, *14* (11), 6414–6417.

(51) Yuan, H.; Liu, X.; Afshinmanesh, F.; Li, W.; Xu, G.; Sun, J.; Lian, B.; Curto, A. G.; Ye, G.; Hikita, Y.; et al. Polarization-Sensitive Broadband Photodetector Using a Black Phosphorus Vertical P-N Junction. *Nat. Nanotechnol.* **2015**, *10* (8), 707–713.

(52) Xia, F.; Wang, H.; Xiao, D.; Dubey, M.; Ramasubramaniam, A. Two-Dimensional Material Nanophotonics. *Nat. Photonics* **2014**, *8* (12), 899–907.

(53) Du, Y.; Liu, H.; Deng, Y.; Ye, P. D. Device Perspective for Black Phosphorus Field-Effect Transistors: Contact Resistance, Ambipolar Behavior, and Scaling. *ACS Nano* **2014**, *8* (10), 10035–10042.

(54) Zhu, W.; Yogeesh, M. N.; Yang, S.; Aldave, S. H.; Kim, J.; Sonde, S. S.; Tao, L.; Lu, N.; Akinwande, D. Flexible Black Phosphorus Ambipolar Transistors, Circuits and AM Demodulator. *Nano Lett.* **2015**, *15* (3), 1883–1890.

(55) Park, J.; Jo, S. B.; Yu, Y.-J.; Kim, Y.; Yang, J. W.; Lee, W. H.; Kim, H. H.; Hong, B. H.; Kim, P.; Cho, K.; et al. Single-Gate Bandgap Opening of Bilayer Graphene by Dual Molecular Doping. *Adv. Mater.* **2012**, *24* (3), 407–411.

(56)  Szafranek, B. N.; Schall, D.; Otto, M.; Neumaier, D.; Kurz, H. High On/off Ratios in Bilayer Graphene Field Effect Transistors Realized by Surface Dopants. *Nano Lett.* **2011**, *11* (7), 2640–2643.

(57)  Kedzierski, J.; Hsu, P.-L.; Healey, P.; Wyatt, P. W.; Keast, C. L.; Sprinkle, M.; Berger, C.; de Heer, W. A. Epitaxial Graphene Transistors on SiC Substrates. *IEEE Trans. Electron Devices* **2008**, *55* (8), 2078–2085.

(58)  Das, A.; Pisana, S.; Chakraborty, B.; Piscanec, S.; Saha, S. K.; Waghmare, U. V; Novoselov, K. S.; Krishnamurthy, H. R.; Geim, A. K.; Ferrari, A. C.; et al. Monitoring Dopants by Raman Scattering in an Electrochemically Top-Gated Graphene Transistor. *Nat. Nanotechnol.* **2008**, *3* (4), 210–215.

(59)  Lemme, M. C.; Echtermeyer, T. J.; Baus, M.; Kurz, H. A Graphene Field-Effect Device. *IEEE Electron Device Lett.* **2007**, *28* (4), 282–284.

(60)  Liang, X.; Fu, Z.; Chou, S. Y. Graphene Transistors Fabricated via Transfer-Printing In Device Active-Areas on Large Wafer. *Nano Lett.* **2007**, *7* (12), 3840–3844.

(61)  Ayari, A.; Cobas, E.; Ogundadegbe, O.; Fuhrer, M. S. Realization and Electrical Characterization of Ultrathin Crystals of Layered Transition-Metal Dichalcogenides. *J. Appl. Phys.* **2007**, *101* (1), 014507.

(62)  Fang, H.; Chuang, S.; Chang, T. C.; Takei, K.; Takahashi, T.; Javey, A. High-Performance Single Layered WSe$_2$ P-FETs with Chemically Doped Contacts. *Nano Lett.* **2012**, *12* (7), 3788–3792.

(63)  Kim, S.; Konar, A.; Hwang, W.-S.; Lee, J. H.; Lee, J.; Yang, J.; Jung, C.; Kim, H.; Yoo, J.-B.; Choi, J.-Y.; et al. High-Mobility and Low-Power Thin-Film Transistors Based on Multilayer MoS2 Crystals. *Nat. Commun.* **2012**, *3*, 1011.

(64)  Wu, W.; De, D.; Chang, S.-C.; Wang, Y.; Peng, H.; Bao, J.; Pei, S.-S. High Mobility and High On/off Ratio Field-Effect Transistors Based on Chemical Vapor Deposited Single-Crystal MoS2 Grains. *Appl. Phys. Lett.* **2013**, *102* (14), 142106.

(65)  Jariwala, D.; Sangwan, V. K.; Late, D. J.; Johns, J. E.; Dravid, V. P.; Marks, T. J.; Lauhon, L. J.; Hersam, M. C. Band-like Transport in High Mobility Unencapsulated Single-Layer MoS Transistors. *Appl. Phys. Lett.* **2013**, *102* (17), 173107.

(66)  Bao, W.; Cai, X.; Kim, D.; Sridhara, K.; Fuhrer, M. S. High Mobility Ambipolar MoS2 Field-Effect Transistors: Substrate and Dielectric Effects. *Appl. Phys. Lett.* **2013**, *102* (4), 042104.

(67)  Liu, W.; Kang, J.; Sarkar, D.; Khatami, Y.; Jena, D.; Banerjee, K. Role of Metal Contacts in Designing High-Performance Monolayer N-Type WSe2 Field Effect Transistors. *Nano Lett.* **2013**, *13* (5), 1983–1990.

(68)  Das, S.; Chen, H.-Y.; Penumatcha, A. V.; Appenzeller, J. High Performance Multilayer MoS2 Transistors with Scandium Contacts. *Nano Lett.* **2013**, *13* (1), 100–105.

(69)  Na, J.; Lee, Y. T.; Lim, J. A.; Hwang, D. K.; Kim, G.-T.; Choi, W. K.; Song, Y.-W. Few-Layer Black Phosphorus Field-Effect Transistors with Reduced Current Fluctuation. *ACS Nano* **2014**, *8* (11), 11753–11762.

(70)  Avsar, A.; Vera-Marun, I. J.; Tan, J. Y.; Watanabe, K.; Taniguchi, T.; Castro Neto, A. H.; Özyilmaz, B. Air-Stable Transport in Graphene-Contacted, Fully Encapsulated Ultrathin Black Phosphorus-Based Field-Effect Transistors. *ACS Nano* **2015**, *9* (4), 4138–4145.

(71)  Kamalakar, M. V.; Madhushankar, B. N.; Dankert, A.; Dash, S. P. Low Schottky Barrier Black Phosphorus Field-Effect Devices with Ferromagnetic Tunnel Contacts. *Small* **2015**, *11* (18), 2209–2216.

(72)  Wang, H.; Wang, X.; Xia, F.; Wang, L.; Jiang, H.; Xia, Q.; Chin, M. L.; Dubey, M.; Han, S. Black Phosphorus Radio-Frequency Transistors. *Nano Lett.* **2014**, *14* (11), 6424–6429.

(73)  Doganov, R. A.; O'Farrell, E. C. T.; Koenig, S. P.; Yeo, Y.; Ziletti, A.; Carvalho, A.; Campbell, D. K.; Coker, D. F.; Watanabe, K.; Taniguchi, T.; et al. Transport Properties of Pristine Few-Layer Black Phosphorus by van Der Waals Passivation in an Inert Atmosphere. *Nat. Commun.* **2015**, *6*, 6647.

(74)  Ling, X.; Wang, H.; Huang, S.; Xia, F.; Dresselhaus, M. S. The Renaissance of Black Phosphorus. *Proc. Natl. Acad. Sci.* **2015**, *112* (15), 201416581.

(75)  Xia, F.; Mueller, T.; Lin, Y.-M.; Valdes-Garcia, A.; Avouris, P. Ultrafast Graphene Photodetector. *Nat. Nanotechnol.* **2009**, *4* (12), 839–843.

(76)  Mueller, T.; Xia, F.; Avouris, P. Graphene Photodetectors for High-Speed Optical Communications. *Nat. Photonics* **2010**, *4* (5), 297–301.

(77)  Gan, X.; Shiue, R.-J.; Gao, Y.; Meric, I.; Heinz, T. F.; Shepard, K.; Hone, J.; Assefa, S.; Englund, D. Chip-Integrated Ultrafast Graphene Photodetector with High Responsivity. *Nat. Photonics* **2013**, *7* (11), 883–887.

(78)  Pospischil, A.; Humer, M.; Furchi, M. M.; Bachmann, D.; Guider, R.; Fromherz, T.; Mueller, T. CMOS-Compatible Graphene Photodetector Covering All Optical Communication Bands. *Nat. Photonics* **2013**, *7* (11), 892–896.

(79)   Wang, X.; Cheng, Z.; Xu, K.; Tsang, H. K.; Xu, J.-B. High-Responsivity Graphene/silicon-Heterostructure Waveguide Photodetectors. *Nat. Photonics* **2013**, *7* (11), 888–891.

(80)   Yin, Z.; Li, H.; Li, H.; Jiang, L.; Shi, Y.; Sun, Y.; Lu, G.; Zhang, Q.; Chen, X.; Zhang, H. Single-Layer MoS2 Phototransistors. *ACS Nano* **2012**, *6* (1), 74–80.

(81)   Furchi, M. M.; Polyushkin, D. K.; Pospischil, A.; Mueller, T. Mechanisms of Photoconductivity in Atomically Thin MoS2. *Nano Lett.* **2014**, *14* (11), 6165–6170.

(82)   Lopez-Sanchez, O.; Lembke, D.; Kayci, M.; Radenovic, A.; Kis, A. Ultrasensitive Photodetectors Based on Monolayer MoS2. *Nat. Nanotechnol.* **2013**, *8* (7), 497–501.

(83)   Choi, W.; Cho, M. Y.; Konar, A.; Lee, J. H.; Cha, G.-B.; Hong, S. C.; Kim, S.; Kim, J.; Jena, D.; Joo, J.; et al. High-Detectivity Multilayer MoS(2) Phototransistors with Spectral Response from Ultraviolet to Infrared. *Adv. Mater.* **2012**, *24* (43), 5832–5836.

(84)   Tsai, D.-S.; Liu, K.-K.; Lien, D.-H.; Tsai, M.-L.; Kang, C.-F.; Lin, C.-A.; Li, L.-J.; He, J.-H. Few-Layer MoS2 with High Broadband Photogain and Fast Optical Switching for Use in Harsh Environments. *ACS Nano* **2013**, *7* (5), 3905–3911.

(85)   Zhang, W.; Huang, J.-K.; Chen, C.-H.; Chang, Y.-H.; Cheng, Y.-J.; Li, L.-J. High-Gain Phototransistors Based on a CVD MoS₂ Monolayer. *Adv. Mater.* **2013**, *25* (25), 3456–3461.

(86)   Perea-López, N.; Lin, Z.; Pradhan, N. R.; Iñiguez-Rábago, A.; Laura Elías, A.; McCreary, A.; Lou, J.; Ajayan, P. M.; Terrones, H.; Balicas, L.; et al. CVD-Grown Monolayered MoS 2 as an Effective Photosensor Operating at Low-Voltage. *2D Mater.* **2014**, *1* (1), 011004.

(87)   Chang, Y.-H.; Zhang, W.; Zhu, Y.; Han, Y.; Pu, J.; Chang, J.-K.; Hsu, W.-T.; Huang, J.-K.; Hsu, C.-L.; Chiu, M.-H.; et al. Monolayer MoSe2 Grown by Chemical Vapor Deposition for Fast Photodetection. *ACS Nano* **2014**, *8* (8), 8582–8590.

(88)   Xia, J.; Huang, X.; Liu, L.-Z.; Wang, M.; Wang, L.; Huang, B.; Zhu, D.-D.; Li, J.-J.; Gu, C.-Z.; Meng, X.-M. CVD Synthesis of Large-Area, Highly Crystalline MoSe2 Atomic Layers on Diverse Substrates and Application to Photodetectors. *Nanoscale* **2014**, *6* (15), 8949–8955.

(89)   Abderrahmane, A.; Ko, P. J.; Thu, T. V; Ishizawa, S.; Takamura, T.; Sandhu, A. High Photosensitivity Few-Layered MoSe2 Back-Gated Field-Effect Phototransistors. *Nanotechnology* **2014**, *25* (36), 365202.

(90) Perea-López, N.; Elías, A. L.; Berkdemir, A.; Castro-Beltran, A.; Gutiérrez, H. R.; Feng, S.; Lv, R.; Hayashi, T.; López-Urías, F.; Ghosh, S.; et al. Photosensor Device Based on Few-Layered WS 2 Films. *Adv. Funct. Mater.* **2013**, *23* (44), 5511–5517.

(91) Huo, N.; Yang, S.; Wei, Z.; Li, S.-S.; Xia, J.-B.; Li, J. Photoresponsive and Gas Sensing Field-Effect Transistors Based on Multilayer WS$_2$ Nanoflakes. *Sci. Rep.* **2014**, *4*, 5209.

(92) Zhang, W.; Chiu, M.-H.; Chen, C.-H.; Chen, W.; Li, L.-J.; Wee, A. T. S. Role of Metal Contacts in High-Performance Phototransistors Based on WSe2 Monolayers. *ACS Nano* **2014**, *8* (8), 8653–8661.

(93) Zhang, J.; Liu, H. J.; Cheng, L.; Wei, J.; Liang, J. H.; Fan, D. D.; Shi, J.; Tang, X. F.; Zhang, Q. J. Phosphorene Nanoribbon as a Promising Candidate for Thermoelectric Applications. *Sci. Rep.* **2014**, *4*, 6452.

(94) Lv, H. Y.; Lu, W. J.; Shao, D. F.; Sun, Y. P. Enhanced Thermoelectric Performance of Phosphorene by Strain-Induced Band Convergence. *Phys. Rev. B* **2014**, *90* (8), 085433.

(95) Lv, H. Y.; Lu, W. J.; Shao, D. F.; Sun, Y. P. Large Thermoelectric Power Factors in Black Phosphorus and Phosphorene. *arXiv* **2014**, 1404.5171.

(96) Qin, G.; Qin, Z.; Yue, S.-Y.; Cui, H.-J.; Zheng, Q.-R.; Yan, Q.-B.; Su, G. Strain Enhanced Anisotropic Thermoelectric Performance of Black Phosphorus. *arXiv* **2014**, 1406.0261.

(97) Qin, G.; Yan, Q.-B.; Qin, Z.; Yue, S.-Y.; Cui, H.-J.; Zheng, Q.-R.; Su, G. Hinge-like Structure Induced Unusual Properties of Black Phosphorus and New Strategies to Improve the Thermoelectric Performance. *Sci. Rep.* **2014**, *4*, 6946.

(98) Flores, E.; Ares, J. R.; Castellanos-Gomez, A.; Barawi, M.; Ferrer, I. J.; Sánchez, C. Thermoelectric Power of Bulk Black-Phosphorus. *Appl. Phys. Lett.* **2015**, *106* (2), 022102.

(99) Fei, R.; Faghaninia, A.; Soklaski, R.; Yan, J.-A.; Lo, C.; Yang, L. Enhanced Thermoelectric Efficiency via Orthogonal Electrical and Thermal Conductances in Phosphorene. *Nano Lett.* **2014**, *14* (11), 6393–6399.

(100) Nam, S.-G.; Ki, D.-K.; Lee, H.-J. Thermoelectric Transport of Massive Dirac Fermions in Bilayer Graphene. *Phys. Rev. B* **2010**, *82* (24), 245416.

(101) Wu, J.; Schmidt, H.; Amara, K. K.; Xu, X.; Eda, G.; Özyilmaz, B. Large Thermoelectricity via Variable Range Hopping in Chemical Vapor Deposition Grown Single-Layer MoS2. *Nano Lett.* **2014**, *14* (5), 2730–2734.

(102) Dobusch, L.; Furchi, M. M.; Pospischil, A.; Mueller, T.; Bertagnolli, E.; Lugstein, A. Electric Field Modulation of Thermovoltage in Single-Layer MoS2. *Appl. Phys. Lett.* **2014**, *105* (25), 253103.

(103) Buscema, M.; Barkelid, M.; Zwiller, V.; van der Zant, H. S. J.; Steele, G. A.; Castellanos-Gomez, A. Large and Tunable Photothermoelectric Effect in Single-Layer MoS2. *Nano Lett.* **2013**, *13* (2), 358–363.

(104) Rodin, A. S.; Carvalho, A.; Castro Neto, A. H. Excitons in Anisotropic Two-Dimensional Semiconducting Crystals. *Phys. Rev. B* **2014**, *90* (7), 075429.

(105) Ong, Z.-Y.; Cai, Y.; Zhang, G.; Zhang, Y.-W. Strong Thermal Transport Anisotropy and Strain Modulation in Single-Layer Phosphorene. *J. Phys. Chem. C* **2014**, *118* (43), 25272–25277.

(106) Fei, R.; Yang, L. Strain-Engineering the Anisotropic Electrical Conductance of Few-Layer Black Phosphorus. *Nano Lett.* **2014**, *14* (5), 2884–2889.

(107) Qin, G.; Yan, Q.-B.; Qin, Z.; Yue, S.-Y.; Hu, M.; Su, G. Anisotropic Intrinsic Lattice Thermal Conductivity of Phosphorene from First Principles. *Phys. Chem. Chem. Phys.* **2015**, *17*, 4854.

(108) Schuster, R.; Trinckauf, J.; Knupfer, M.; Büchner, B. Anisotropic Particle-Hole Excitations in Black Phosphorus. *Phys. Rev. Lett.* **2015**, *115*, 026404.

(109) Jiang, J.-W. The Third Principle Direction Besides Armchair and Zigzag in Single-Layer Black Phosphorus. *Nanotechnology* **2015**, *26* (36), 365702.

(110) Jiang, J.-W. Thermal Conduction in Single-Layer Black Phosphorus: Highly Anisotropic? *Nanotechnology* **2015**, *26* (5), 055701.

(111) Popović, Z. S.; Kurdestany, J. M.; Satpathy, S. Electronic Structure and Anisotropic Rashba Spin-Orbit Coupling in Monolayer Black Phosphorus. *Phys. Rev. B* **2015**, *92* (3), 035135.

(112) He, J.; He, D.; Wang, Y.; Cui, Q.; Bellus, M. Z.; Chiu, H.-Y.; Zhao, H. Exceptional and Anisotropic Transport Properties of Photocarriers in Black Phosphorus. *ACS Nano* **2015**, *9* (6), 6436–6442.

(113) Lu, W.; Ma, X.; Fei, Z.; Zhou, J.; Zhang, Z.; Jin, C.; Zhang, Z. Probing the Anisotropic Behaviors of Black Phosphorus by Transmission Electron Microscopy, Angular-Dependent Raman Spectra, and Electronic Transport Measurements. *Appl. Phys. Lett.* **2015**, *107* (2), 021906.

(114) Luo, Z.; Maassen, J.; Deng, Y.; Du, Y.; Lundstrom, M. S.; Ye, P. D.; Xu, X. Anisotropic in-Plane Thermal Conductivity Observed in Few-Layer Black Phosphorus. *arXiv* **2015**, 1503.06167.

(115) Ge, S.; Li, C.; Zhang, Z.; Zhang, C.; Zhang, Y.; Qiu, J.; Wang, Q.; Liu, J.; Jia, S.; Feng, J.; et al. Dynamical Evolution of Anisotropic Response in Black Phosphorus under Ultrafast Photoexcitation. *Nano Lett.* **2015**, *15* (7), 4650–4656.

(116) Chaves, A.; Low, T.; Avouris, P.; Çakır, D.; Peeters, F. M. Anisotropic Exciton Stark Shift in Black Phosphorus. *Phys. Rev. B* **2015**, *91* (15), 155311.

(117) Zhang, S.; Yang, J.; Xu, R.; Wang, F.; Li, W.; Ghufran, M.; Zhang, Y.-W.; Yu, Z.; Zhang, G.; Qin, Q.; et al. Extraordinary Photoluminescence and Strong Temperature/angle-Dependent Raman Responses in Few-Layer Phosphorene. *ACS Nano* **2014**, *8* (9), 9590–9596.

(118) Jiang, J.-W.; Park, H. S. Mechanical Properties of Single-Layer Black Phosphorus. *J. Phys. D. Appl. Phys.* **2014**, *47* (38), 385304.

(119) Li, Y.; Yang, S.; Li, J. Modulation of the Electronic Properties of Ultrathin Black Phosphorus by Strain and Electrical Field. *J. Phys. Chem. C* **2014**, *118* (41), 23970–23976.

(120) Ribeiro, H. B.; Pimenta, M. A.; de Matos, C. J. S.; Moreira, R. L.; Rodin, A. S.; Zapata, J. D.; de Souza, E. A. T.; Castro Neto, A. H. Unusual Angular Dependence of the Raman Response in Black Phosphorus. *ACS Nano* **2015**, *9* (4), 4270–4276.

(121) Wu, J.; Mao, N.; Xie, L.; Xu, H.; Zhang, J. Identifying the Crystalline Orientation of Black Phosphorus Using Angle-Resolved Polarized Raman Spectroscopy. *Angew. Chemie* **2015**, *127* (8), 2396–2399.

(122) Wei, Q.; Peng, X. Superior Mechanical Flexibility of Phosphorene and Few-Layer Black Phosphorus. *Appl. Phys. Lett.* **2014**, *104* (25), 251915.

(123) Cakir, D.; Sevik, C.; Peeters, F. M. Remarkable Effect of Stacking on the Electronic and Optical Properties of Few Layer Black Phosphorus. *arXiv* **2015**, 1506.04707.

(124) Low, T.; Roldán, R.; Wang, H.; Xia, F.; Avouris, P.; Moreno, L. M.; Guinea, F. Plasmons and Screening in Monolayer and Multilayer Black Phosphorus. *Phys. Rev. Lett.* **2014**, *113* (10), 106802.

(125) Appalakondaiah, S.; Vaitheeswaran, G.; Lebègue, S.; Christensen, N. E.; Svane, A. Effect of van Der Waals Interactions on the Structural and Elastic Properties of Black Phosphorus. *Phys. Rev. B* **2012**, *86* (3), 035105.

(126) Wang, Z.; Feng, P. X.-L. Design of Black Phosphorus 2D Nanomechanical Resonators by Exploiting the Intrinsic Mechanical Anisotropy. *2D Mater.* **2015**, *2* (2), 021001.

(127) Yasaei, P.; Kumar, B.; Foroozan, T.; Wang, C.; Asadi, M.; Tuschel, D.; Indacochea, J. E.; Klie, R. F.; Salehi-Khojin, A. High-Quality Black Phosphorus Atomic Layers by Liquid-Phase Exfoliation. *Adv. Mater.* **2015**, *27* (11), 1887–1892.

(128) Hanlon, D.; Backes, C.; Doherty, E.; Cucinotta, C. S.; Berner, N. C.; Boland, C.; Lee, K.; Lynch, P.; Gholamvand, Z.; Harvey, A.; et al. Liquid Exfoliation of Solvent-Stabilised Black Phosphorus: Applications beyond Electronics. *arXiv* **2015**, 1501.01881.

(129) Sresht, V.; Pádua, A. A. H.; Blankschtein, D. Liquid-Phase Exfoliation of Phosphorene: Design Rules from Molecular Dynamics Simulations. *ACS Nano* **2015**, *9* (8), 8255–8268.

(130) Kang, J.; Wood, J. D.; Wells, S. A.; Lee, J.-H.; Liu, X.; Chen, K.-S.; Hersam, M. C. Solvent Exfoliation of Electronic-Grade, Two-Dimensional Black Phosphorus. *ACS Nano* **2015**, *9* (4), 3596–3604.

(131) Luo, Z.-C.; Liu, M.; Guo, Z.-N.; Jiang, X.-F.; Luo, A.-P.; Zhao, C.-J.; Yu, X.-F.; Xu, W.-C.; Zhang, H. Microfiber-Based Few-Layer Black Phosphorus Saturable Absorber for Ultra-Fast Fiber Laser. *arXiv* **2015**, 1505.03035.

(132) Jiang, T.; Yin, K.; Zheng, X.; Yu, H.; Cheng, X.-A. Black Phosphorus as a New Broadband Saturable Absorber for Infrared Passively Q-Switched Fiber Lasers. *arXiv* **2015**, 1504.07341.

(133) Yang, Z.; Hao, J.; Yuan, S.; Lin, S.; Yau, H. M.; Dai, J.; Lau, S. P. Field-Effect Transistors Based on Amorphous Black Phosphorus Ultrathin Films by Pulsed Laser Deposition. *Adv. Mater.* **2015**, *27* (25), 3748–3754.

(134) Li, X.; Cai, W.; An, J.; Kim, S.; Nah, J.; Yang, D.; Piner, R.; Velamakanni, A.; Jung, I.; Tutuc, E.; et al. Large-Area Synthesis of High-Quality and Uniform Graphene Films on Copper Foils. *Science* **2009**, *324* (5932), 1312–1314.

(135) Reina, A.; Jia, X.; Ho, J.; Nezich, D.; Son, H.; Bulovic, V.; Dresselhaus, M. S.; Kong, J. Large Area, Few-Layer Graphene Films on Arbitrary Substrates by Chemical Vapor Deposition. *Nano Lett.* **2009**, *9* (1), 30–35.

(136) Van der Zande, A. M.; Huang, P. Y.; Chenet, D. A.; Berkelbach, T. C.; You, Y.; Lee, G.-H.; Heinz, T. F.; Reichman, D. R.; Muller, D. A.; Hone, J. C. Grains and Grain Boundaries in Highly Crystalline Monolayer Molybdenum Disulphide. *Nat. Mater.* **2013**, *12* (6), 554–561.

(137) Najmaei, S.; Liu, Z.; Zhou, W.; Zou, X.; Shi, G.; Lei, S.; Yakobson, B. I.; Idrobo, J.-C.; Ajayan, P. M.; Lou, J. Vapour Phase Growth and Grain Boundary Structure of Molybdenum Disulphide Atomic Layers. *Nat. Mater.* **2013**, *12* (8), 754–759.

(138) Island, J. O.; Steele, G. A.; Zant, H. S. J. van der; Castellanos-Gomez, A. Environmental Instability of Few-Layer Black Phosphorus. *2D Mater.* **2015**, *2* (1), 011002.

(139) Wood, J. D.; Wells, S. A.; Jariwala, D.; Chen, K.-S.; Cho, E.; Sangwan, V. K.; Liu, X.; Lauhon, L. J.; Marks, T. J.; Hersam, M. C. Effective Passivation of Exfoliated Black Phosphorus Transistors against Ambient Degradation. *Nano Lett.* **2014**, *14* (12), 6964–6970.

(140) Kim, J.-S.; Liu, Y.; Zhu, W.; Kim, S.; Wu, D.; Tao, L.; Dodabalapur, A.; Lai, K.; Akinwande, D. Toward Air-Stable Multilayer Phosphorene Thin-Films and Transistors. *Sci. Rep.* **2015**, *5*, 8989.

(141) Favron, A.; Gaufrès, E.; Fossard, F.; Phaneuf-L'Heureux, A.-L.; Tang, N. Y.-W.; Lévesque, P. L.; Loiseau, A.; Leonelli, R.; Francoeur, S.; Martel, R. Photooxidation and Quantum Confinement Effects in Exfoliated Black Phosphorus. *Nat. Mater.* **2015**, *14* (8), 826–832.

(142) Kulish, V. V; Malyi, O. I.; Persson, C.; Wu, P. Adsorption of Metal Adatoms on Single-Layer Phosphorene. *Phys. Chem. Chem. Phys.* **2015**, *17* (2), 992–1000.

(143) Ziletti, A.; Carvalho, A.; Campbell, D. K.; Coker, D. F.; Castro Neto, A. H. Oxygen Defects in Phosphorene. *Phys. Rev. Lett.* **2015**, *114* (4), 046801.

(144) Tayari, V.; Hemsworth, N.; Fakih, I.; Favron, A.; Gaufrès, E.; Gervais, G.; Martel, R.; Szkopek, T. Two-Dimensional Magnetotransport in a Black Phosphorus Naked Quantum Well. *Nat. Commun.* **2015**, *6*, 7702.

(145) Gillgren, N.; Wickramaratne, D.; Shi, Y.; Espiritu, T.; Yang, J.; Hu, J.; Wei, J.; Liu, X.; Mao, Z.; Watanabe, K.; et al. Gate Tunable Quantum Oscillations in Air-Stable and High Mobility Few-Layer Phosphorene Heterostructures. *2D Mater.* **2014**, *2* (1), 011001.

(146) Li, L.; Ye, G. J.; Tran, V.; Fei, R.; Chen, G.; Wang, H.; Wang, J.; Watanabe, K.; Taniguchi, T.; Yang, L.; et al. Quantum Oscillations in a Two-Dimensional Electron Gas in Black Phosphorus Thin Films. *Nat. Nanotechnol.* **2015**, *10* (7), 608–613.

(147) Cao, Y.; Mishchenko, A.; Yu, G. L.; Khestanova, K.; Rooney, A. P.; Prestat, E.; Kretinin, A. V.; Blake, P.; Shalom, M. B.; Balakrishnan, G.; et al. Quality Heterostructures from Two Dimensional Crystals Unstable in Air by Their Assembly in Inert Atmosphere. *Nano Lett.* **2015**, *15* (8), 4914–4921.

(148) Chen, X.; Wu, Y.; Wu, Z.; Han, Y.; Xu, S.; Wang, L.; Ye, W.; Han, T.; He, Y.; Cai, Y.; et al. High-Quality Sandwiched Black Phosphorus Heterostructure and Its Quantum Oscillations. *Nat. Commun.* **2015**, *6*, 7315.

(149) Li, L.; Yang, F.; Ye, G. J.; Zhang, Z.; Zhu, Z.; Lou, W.-K.; Li, L.; Watanabe, K.; Taniguchi, T.; Chang, K.; et al. Quantum Hall Effect in Black Phosphorus Two-Dimensional Electron Gas. *arXiv* **2015**, 1504.07155.

(150) Wang, L.; Meric, I.; Huang, P. Y.; Gao, Q.; Gao, Y.; Tran, H.; Taniguchi, T.; Watanabe, K.; Campos, L. M.; Muller, D. A.; et al. One-Dimensional Electrical Contact to a Two-Dimensional Material. *Science* **2013**, *342* (6158), 614–617.

(151) Cui, X.; Lee, G.-H.; Kim, Y. D.; Arefe, G.; Huang, P. Y.; Lee, C.-H.; Chenet, D. A.; Zhang, X.; Wang, L.; Ye, F.; et al. Multi-Terminal Transport Measurements of MoS2 Using a van Der Waals Heterostructure Device Platform. *Nat. Nanotechnol.* **2015**, *10* (6), 534–540.

(152) Luo, X.; Lu, X.; Koon, G. K. W.; Castro Neto, A. H.; Özyilmaz, B.; Xiong, Q.; Quek, S. Y. Large Frequency Change with Thickness in Interlayer Breathing Mode-Significant Interlayer Interactions in Few Layer Black Phosphorus. *Nano Lett.* **2015**, *15* (6), 3931–3938.

(153) Ling, X.; Liang, L.; Huang, S.; Puretzky, A. A.; Geohegan, D. B.; Sumpter, B. G.; Kong, J.; Meunier, V.; Dresselhaus, M. S. Low-Frequency Interlayer Breathing Modes in Few-Layer Black Phosphorus. *Nano Lett.* **2015**, *15* (6), 4080–4088.

(154) Dong, S.; Zhang, A.; Liu, K.; Ji, J.; Ye, Y. G.; Luo, X. G.; Chen, X. H.; Ma, X.; Jie, Y.; Chen, C.; et al. Ultralow-Frequency Collective Compression Mode and Strong Interlayer Coupling in Multilayer Black Phosphorus. *arXiv* **2015**, 1503.06577.

(155) Yang, J.; Xu, R.; Pei, J.; Myint, Y. W.; Wang, F.; Wang, Z.; Zhang, S.; Yu, Z.; Lu, Y. Unambiguous Identification of Monolayer Phosphorene by Phase-Shifting Interferometry. *arXiv* **2014**, 1412.6701.

(156) Edmonds, M. T.; Tadich, A.; Carvalho, A.; Ziletti, A.; O'Donnell, K. M.; Koenig, S. P.; Coker, D. F.; Özyilmaz, B.; Neto, A. H. C.; Fuhrer, M. S. Creating a Stable Oxide at the Surface of Black Phosphorus. *ACS Appl. Mater. Interfaces* **2015**, *7* (27), 14557–14562.

(157) Han, C. Q.; Yao, M. Y.; Bai, X. X.; Miao, L.; Zhu, F.; Guan, D. D.; Wang, S.; Gao, C. L.; Liu, C.; Qian, D.; et al. Electronic Structure of Black Phosphorus Studied by Angle-Resolved Photoemission Spectroscopy. *Phys. Rev. B* **2014**, *90* (8), 085101.

(158) Xiang, D.; Han, C.; Wu, J.; Zhong, S.; Liu, Y.; Lin, J.; Zhang, X.-A.; Ping Hu, W.; Özyilmaz, B.; Neto, A. H. C.; et al. Surface Transfer Doping Induced Effective

Modulation on Ambipolar Characteristics of Few-Layer Black Phosphorus. *Nat. Commun.* **2015**, *6*, 6485.

(159) Padilha, J. E.; Fazzio, A.; da Silva, A. J. R. Van Der Waals Heterostructure of Phosphorene and Graphene: Tuning the Schottky Barrier and Doping by Electrostatic Gating. *Phys. Rev. Lett.* **2015**, *114* (6), 066803.

(160) Chen, P.; Xiang, J.; Yu, H.; zhang, J.; Xie, G.; Wu, S.; Lu, X.; Wang, G.; Zhao, J.; Wen, F.; et al. Gate Tunable MoS 2 –black Phosphorus Heterojunction Devices. *2D Mater.* **2015**, *2* (3), 034009.

(161) Gehring, P.; Urcuyo, R.; Duong, D. L.; Burghard, M.; Kern, K. Thin-Layer Black phosphorous/GaAs Heterojunction P-N Diodes. *Appl. Phys. Lett.* **2015**, *106* (23), 233110.

(162) Yan, R.; Fathipour, S.; Han, Y.; Song, B.; Xiao, S.; Li, M.; Ma, N.; Protasenko, V.; Muller, D. A.; Jena, D.; et al. Esaki Diodes in van Der Waals Heterojunctions with Broken-Gap Energy Band Alignment. *Nano Lett.* **2015**, *15* (9), 5791–5798.

(163) Konstantatos, G.; Badioli, M.; Gaudreau, L.; Osmond, J.; Bernechea, M.; Garcia de Arquer, F. P.; Gatti, F.; Koppens, F. H. L. Hybrid Graphene-Quantum Dot Phototransistors with Ultrahigh Gain. *Nat. Nanotechnol.* **2012**, *7* (6), 363–368.

(164) Sundaram, R. S.; Engel, M.; Lombardo, A.; Krupke, R.; Ferrari, A. C.; Avouris, P.; Steiner, M. Electroluminescence in Single Layer MoS2. *Nano Lett.* **2013**, *13* (4), 1416–1421.

(165) Jo, S.; Ubrig, N.; Berger, H.; Kuzmenko, A. B.; Morpurgo, A. F. Mono- and Bilayer WS2 Light-Emitting Transistors. *Nano Lett.* **2014**, *14* (4), 2019–2025.

(166) Cheng, R.; Li, D.; Zhou, H.; Wang, C.; Yin, A.; Jiang, S.; Liu, Y.; Chen, Y.; Huang, Y.; Duan, X. Electroluminescence and Photocurrent Generation from Atomically Sharp WSe2/MoS2 Heterojunction P-N Diodes. *Nano Lett.* **2014**, *14* (10), 5590–5597.

(167) Roldán, R.; Castellanos-Gomez, A.; Cappelluti, E.; Guinea, F. Strain Engineering in Semiconducting Two-Dimensional Crystals. *J. Phys. Condens. Matter* **2015**, *27* (31), 313201.

(168) He, K.; Poole, C.; Mak, K. F.; Shan, J. Experimental Demonstration of Continuous Electronic Structure Tuning via Strain in Atomically Thin MoS2. *Nano Lett.* **2013**, *13* (6), 2931–2936.

(169) Conley, H. J.; Wang, B.; Ziegler, J. I.; Haglund, R. F.; Pantelides, S. T.; Bolotin, K. I. Bandgap Engineering of Strained Monolayer and Bilayer MoS2. *Nano Lett.* **2013**, *13* (8), 3626–3630.

(170)  Desai, S. B.; Seol, G.; Kang, J. S.; Fang, H.; Battaglia, C.; Kapadia, R.; Ager, J. W.; Guo, J.; Javey, A. Strain-Induced Indirect to Direct Bandgap Transition in Multilayer WSe 2. *Nano Lett.* **2014**, *14* (8), 4592–4597.

(171)  Scalise, E.; Houssa, M.; Pourtois, G.; Afanas'ev, V.; Stesmans, A. Strain-Induced Semiconductor to Metal Transition in the Two-Dimensional Honeycomb Structure of MoS2. *Nano Res.* **2011**, *5* (1), 43–48.

(172)  Castellanos-Gomez, A.; Roldán, R.; Cappelluti, E.; Buscema, M.; Guinea, F.; van der Zant, H. S. J.; Steele, G. A. Local Strain Engineering in Atomically Thin MoS2. *Nano Lett.* **2013**, *13* (11), 5361–5366.

(173)  Hui, Y. Y.; Liu, X.; Jie, W.; Chan, N. Y.; Hao, J.; Hsu, Y.-T.; Li, L.-J.; Guo, W.; Lau, S. P. Exceptional Tunability of Band Energy in a Compressively Strained Trilayer MoS2 Sheet. *ACS Nano* **2013**, *7* (8), 7126–7131.

(174)  Sorkin, V.; Zhang, Y. W. The Deformation and Failure Behaviour of Phosphorene Nanoribbons under Uniaxial Tensile Strain. *2D Mater.* **2015**, *2* (3), 035007.

(175)  Rodin, A. S.; Carvalho, A.; Castro Neto, A. H. Strain-Induced Gap Modification in Black Phosphorus. *Phys. Rev. Lett.* **2014**, *112* (17), 176801.

(176)  Elahi, M.; Khaliji, K.; Tabatabaei, S. M.; Pourfath, M.; Asgari, R. Modulation of Electronic and Mechanical Properties of Phosphorene through Strain. *Phys. Rev. B* **2015**, *91* (11), 115412.

(177)  Çakır, D.; Sahin, H.; Peeters, F. M. Tuning of the Electronic and Optical Properties of Single-Layer Black Phosphorus by Strain. *Phys. Rev. B* **2014**, *90* (20), 205421.

(178)  Quereda, J.; Parente, V.; San-José, P.; Agraït, N.; Rubio-Bollinger, G.; Guinea, F.; Roldán, R.; Castellanos-Gomez, A. Quantum Confinement in Black Phosphorus through Strain-Engineered Rippling. *arXiv* **2015**, 1509.01182.

(179)  Li, X.; Deng, B.; Wang, X.; Chen, S.; Vaisman, M.; Karato, S.; Pan, G.; Larry Lee, M.; Cha, J.; Wang, H.; et al. Synthesis of Thin-Film Black Phosphorus on a Flexible Substrate. *2D Mater.* **2015**, *2* (3), 031002.